\documentclass [12pt,twoside]{article}           % version LATEX2E
\usepackage[left,modulo]{lineno}
\usepackage{setspace}
\countdef\arxiv=201
\ifnum\arxiv=0 \input cpreb.tex \fi

\ifnum\arxiv=1

\usepackage{epsfig,times,lscape}
\usepackage[usenames]{color}

    \ifnum\count102=1

    \fi

    \ifnum\count102=2
\fi

\newcommand{\nc}{\newcommand}

     \ifnum\count101=1
\nc{\qI}[1]{\section{{#1}}}
\nc{\qA}[1]{\subsection{{#1}}}
\nc{\qun}[1]{\subsubsection{{#1}}}
\nc{\qa}[1]{\paragraph{{#1}}}

\def\qpar{\vskip 2mm plus 0.2mm minus 0.2mm}
\def\qL{\hfill \break}
     \fi 

      \ifnum\count101=2
 \nc{\qI}[1]{\parindent=0mm \vskip 8mm 
{\centerline{\LARGE \color{red}#1}}\vskip 3mm}
\nc{\qA}[1]{\vskip 2.5mm \noindent 
{{\bf\large\color{blue}  #1}} \vskip 1mm \parindent=0mm}
 \nc{\qun}[1]{\vskip 1mm \noindent {\sl #1 }\quad }

\def\qL{\hfill \break}
\def\qpar{\vskip 2mm plus 0.2mm minus 0.2mm}

      \fi
\def\qth{\vrule height 12pt depth 0pt width 0pt}
\def\qtb{\vrule height 0pt depth 5pt width 0pt}

\nc{\qfoot}[1]{\footnote{{#1}}}

      \ifnum\count101=1
\def\qbu{\hfill \par \hskip 6mm $ \bullet $ \hskip 2mm}
\def\qee#1{\hfill \par \hskip 6mm (#1) \hskip 2 mm}
      \fi
      \ifnum\count101=2
\def\qbu{\hfill \par \hskip 4mm $ \bullet $ \hskip 2mm}
\def\qee#1{\hfill \par \hskip 4mm (#1) \hskip 2 mm}
      \fi

\def\qparr{ \vskip 1.0mm plus 0.2mm minus 0.2mm \hangindent=10mm
\hangafter=1}

     \ifnum\count101=1 
 \def\qdec#1{\parindent=0mm\par {\leftskip=2cm {#1} \par}}
     \fi
     \ifnum\count101=2
  \def\qdec#1{\parindent=0mm \par {\leftskip=1cm {#1} \par}}
  
  \def\qcitb#1{\noindent \hbox to 102mm{\hfill \small #1} \vskip 1mm}
      \fi

 \def\qpages#1{\count102=0{\loop\advance\count102 by 1
 \null \vfill\eject \ifnum\count102<#1 \repeat}}

\def\qn#1{\eqno \hbox{(#1)}}

\def\qth{\vrule height 12pt depth 0pt width 0pt}
\def\qtb{\vrule height 0pt depth 5pt width 0pt}

\def\qv{\vskip 0.1mm plus 0.05mm minus 0.05mm}
\def\qhu{\hskip 0.6mm}
\def\qhv{\hskip 3mm}

\def\qhw{\hskip 1.5mm}
\def\qleg#1#2#3{\noindent {\bf \small #1\qhw}{\small #2\qhw}{\it \small #3}\qv }
\fi
\begin{document}
\thispagestyle{empty}

% --------------------------------------------------------------------

      % Hauts de pages et numerotation

          % Remarque: sans le \protect --> message d'erreur (ordre fragile)
\markboth{{\sl \hfill  \hfill \protect\phantom{3}}}
        {{\protect\phantom{3}\sl \hfill  \hfill}}

% -------------------------------------------------------------------
\color{yellow} 
%\hrule height 20mm depth 10mm width 170mm 
\hrule height 10mm depth 10mm width 170mm 
\color{black}

\vskip -12mm 

\centerline{\bf \Large Impact of marital status on health}
\vskip 15mm

\centerline{\large 
Peter Richmond$ ^1 $ and Bertrand M. Roehner$ ^2 $
}

\vskip 10mm
\large

{\bf Abstract}\qL
The Farr-Bertillon law states that the 
mortality rate of single and widowed persons 
is about three times the rate of married people
of same age.
This excess mortality can be measured
with good accuracy for all ages except for young
widowers. The reason is that, at least nowadays, 
very few people become widowed under the age of 30.
Here we show that disability data from census records
can also be used as 
a reliable substitute for mortality rates. 
In fact excess-disability and excess-mortality
go hand in hand. Moreover, as
there are
about ten times more cases of disability than deaths,
the disability variable is able to offer 
more accurate measurements in all cases where
the number of deaths is small. This
not only allows a more accurate investigation 
of the young widower effect; it confirms that, as
already suspected from death rate data,
there is a huge spike between the ages of 20 and 30.
\qL
By using disability rates we can also study
additional features not accessible using death rate data.
For example we can examine the
health impact of a change in living place.
The observed temporary inflated disability rate confirms
what could be
expected by invoking the ``Transient Shock''
conjecture formulated by the authors in a previous paper.

\vskip 5mm
\centerline{\it \small Provisional. Version of 5 February 2017. 
Comments are welcome.}
\vskip 5mm

{\small Key-words: marital status, migration, death rate, 
disability}

\vskip 5mm

{\normalsize
1: School of Physics, Trinity College Dublin, Ireland.
Email: peter\_richmond@ymail.com \qL
2: Institute for Theoretical and High Energy Physics (LPTHE),
University Pierre and Marie Curie, Paris, France. 
Email: roehner@lpthe.jussieu.fr
}

\vfill\eject

\qI{Introduction}

The Farr-Bertillon law (Bertillon 1872, Richmond et al. 2016a)
states that married persons have a lower
death rate than non-married persons be they single, divorced 
or widowed.
In its simplest form this law has been known
for over one century. In a previous paper (Richmond et al. 2016)
the present authors added several new features to our knowledge
of this effect.
\qbu As a function of age
the death ratio, namely the ratio (rate of non-married)/(rate of married),
has the same shape today as it had 130 years ago.
\qbu The law holds not only for the global death rate
but also separately for all major classes of diseases,
e.g. heart disease, cerebrovascular diseases, pulmonary
diseases. The law is also valid for suicide and accidental
causes of death.
\qbu Chinese data for 1990 show a pattern that is very similar
to that observed in western countries. These data
provide a more accurate
picture of the young widower effect than that provided by
western data due to:
(i) the sheer size of the Chinese
population and 
(ii) the tradition of early marriage which was still 
common even in 1990 for some rural provinces. It turns out that,
around the age of 20, for men as well as for women,
the widowed/married death ratio displays a huge spike with
amplitude about 20.
\qpar

As a reminder and for the purpose of comparison
with subsequent disability-based graphs, Fig. 1
shows the shape of the death rate ratio for
widowed and single persons.
%
%%-----------------------------------------------
%%%% FIG -> DEATH RATIO: WIDOWED/MARRIED, SINGLE/MARRIED
\begin{figure}[htb]
\centerline{\psfig{width=10cm,figure=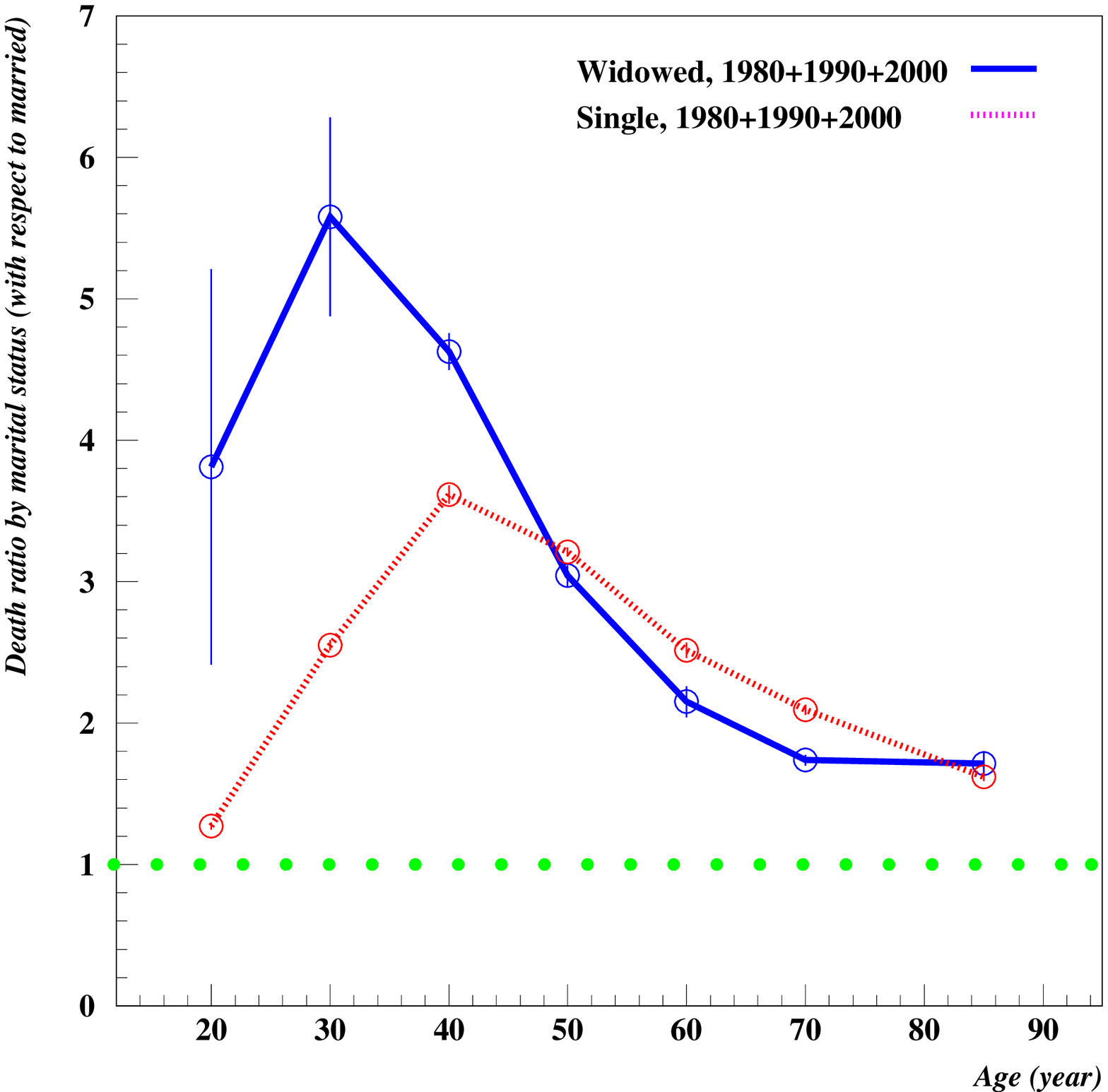}}
\qleg{Fig.\qhu 1\qhv Age-specific death ratio
with respect to married persons in the United States.} 
{The death ratio is the death rate of widowed persons divided
by the death rate of married persons. 
This renormalization removes the exponential behavior
common to both rates. These curves should be compared
with the curves of Fig. 2b which, instead of death, are based
on disability. The data points give the death ratio for
10-year age intervals.
The error bars represent $ \pm \sigma $ where $ \sigma $
is the standard deviation of the average.
This corresponds to
a confidence level of 0.68. The curve for ``single'' is the
average of 15 annual series which is why the error bars
as so small that they are hardly visible.} 
{Source: Richmond et al. (2016): widowed: p.757, 
single: p. 755.}
\end{figure}
%-------------------------------------------------
%
Mortality data are based on death certificates which record
basic information about the deceased and the 
circumstances and causes of the death.
\qpar

In this paper we use census data. This is a
completely different type of statistical in the sense
that mortality data rely on continuous monitoring 
through the vital statistics network whereas
censuses are taken every 10 years
and provide a static picture of the whole population.
It is difficult to decide which one of the two sources
is more reliable but for our present study
what is important is the fact 
that they are independent and very different from 
one another.\qL
One might suspect the 
young widower effect observed from death rate data is a statistical artifact
because under the age of 25 the number of cases
is very small which makes these data very sensitive
to any under-recording. It is therefore
important to observe this effect through a different data 
source.
If census data display the same effect it will
make us more confident that it is indeed genuine.
\qpar

The census data that we use here are disability variables.
Thus, the death ratio will be replaced by a disability
ratio.
A distinct advantage of the census data is that it is possible to compute 
the disability ratio for every year of age.
This contrasts with mortality data which are
given for only 5-year or 10-year age intervals.
This difference becomes particularly
crucial for the age interval around 20
where the young widower effect is expected.
\qpar

The paper proceeds as follows.\qL
In section 2 we describe the disability data and explain our methodology.
Section 3 shows the results of our investigation.
Next we explore the effect of a housing relocation
on the disability rate from which we see that there
is a distinctive increase albeit of a much
smaller magnitude than for the effect of marital
status. Finally, we examine the adverse short-term effect of
marriage.
\qpar

Before we start note that the present paper
follows exactly the methodology associated with physics.
Thus in physics a newly identified phenomenon
is accepted as being real
only after it has been observed in
different experiments and under diverse conditions.
For example, the speed of light was ultimately measured
hnumerous times using various methods and with ever
increasing accuracy. As a result it
has become an unshakable pillar of modern physics.
By way of contrast, what is striking in social science is that conflicting observations
are cited in review papers and apparently accepted
without any real attempt being made 
to discriminate between fact and artifact. 
How can something solid be built on such shaky foundations?
\qL
Our present analysis, based on disability data, of the
incidence of marital status
is akin to setting up a new experiment. If it confirms
and supplements previous observations
that strengthens our confidence in the Farr-Bertillon
law and the young widower effect.

\qI{Data and methodology}

\qA{Disability data}

The US censuses of 1980 and 1990 asked the
following  question:
\qdec{``Does this person have a physical, mental, or other health
  condition which has lasted for 6 or more months and which
prevents this person from working at a job?''}
At first sight it might seem that the question
concerned only persons who are not retired and it is true that
in the census of 1970 the question concerned only persons 
under the age of 65. But in the censuses of 1980 and 1990
the question was asked of every person above the age of 15.
In these cases ``prevents this person from working at a job'' must be understood
as “a condition serious enough to prevent you from working
at a bold{possible} job''. As a matter of fact, the 
data show a steady increase with age
of the proportion of the persons afflicted with a
disability. By the age of 85 the percentage reaches about
50\% (see Fig. 4).
\qpar

The census of 1980 (but not the one of 1990) contained
another question about disability which was the following:
\qdec{``Does this person have a physical, mental, or other health
  condition which has lasted for 6 or more months and which
limits or prevents this person from using public transportation?''}
Naturally one would expect the two disability variables
to be strongly correlated and this is indeed the case. 
In what follows we use only the work disability variable
because the other data is not available for 1990.
\qpar

A consistency test consists in
checking whether the age-specific disability variable is
correlated with the age-specific death rate. This is indeed true not only for
the whole population but also for its subsets. For instance, for
widowed persons, if one denotes the number of persons with a disability
in the 5\% sample of the 1980 census 
by $ d_h $ and the number of deaths in the whole population
by $ D $ one gets%
\qfoot{The death data by age and marital status are from
the 1980 volume of ``Vital Statistics of the United States'', table 1-31.}%
: 
$$ D=1.18 d_h^a,\quad a=1.06\pm0.07 \qn{1} $$

This relationship shows two things:
\qee{1} The two variables are almost proportional to one another.
However, as death numbers are available only for 10-year age intervals
equation (1) indicates a global rather than a year-by-year
proportionality. Actually, on a yearly scale
one would not expect a close connection for a fairly simple reason.
The death numbers are annual variables whereas
the disability numbers are cumulative variables in the sense that
disabilities which last more than one year will be added
together. Thus, in a given year, in addition to the current number
the disability level will also reflect extant
past disabilities.
\qee{2} As $ D $ and $ d_h $ are of same magnitude in populations
of different sizes, we see that
for the entire population 
the number of persons with a disability would be about 20 times larger
than the number of deaths. Therefore, if one could get disability
data for the whole population one would be in an excellent position
to study the young widower effect. Unfortunately,
as explained in the next subsection
only 1\% and 5\% samples are so far available.

\qA{The IPUMS database}
Now that we know the 1980 and 1990 censuses contain
the data we need, how can we access it?
Over the past decades the University of Minnesota has developed
a database containing {\it individual} records of
all US censuses 
except that for 1890 which was destroyed in a fire.
``Individual'' means that the database will deliver files
in which each line corresponds to one person
and contains as many coded variables as the user selects. Access is free and the data are provided in several
formats. For our research we have used the ``text only'' format, formerly called the ASCII (American Standard Code for Information
Interchange) format. However, there are two limitations.
\qbu The data are available only in the form of random
samples, either 1\% or 5\% samples%
\qfoot{Full scale data are available only for a few censuses,
for instance in 1940.}%
. 
The results given below
are based on the 5\% samples of the 1980 and 1990 censuses. As the disability variable exists only above the
age of 15 and under 90 we limited our samples to the
age interval $ (16,1989) $. As a result the files of the
5\% sample
of 1980 and 1990 contained 8,746,006 and 9,529,970 lines
respectively.
\qbu For reasons of confidentiality the variables we use
proposed do not allow precise localization of the individuals.
Information about their place of residence is limited to
county level or even to a cluster of several
counties when the counties are small. As the present
investigation does not use residence location variables
this limitation is of no concern.

\qA{The small $ n $ difficulty}
As already explained one of our main objectives is to explore
the Farr-Bertillon effect for widowed persons
in the age interval from 20 to 30.
The main difficulty comes from the fact that even for a large
country like the United states there are only few young widowers
and among them only a small percentage with a
disability. This difficulty is illustrated in Table 1 which gives
the number of widowed persons and the numbers within this group with a disability.

%%-----------------------------------------------
\begin{table}[htb]

\small

\centerline{\bf Table 1\quad Number of widowed persons 
in the 5\% sample of the US census of 1980}

\vskip 5mm
\hrule
\vskip 0.7mm
\hrule
\vskip 2mm

$$ \matrix{
\qtb
\hbox{Age}  &  \hfill \hbox{Men} & \hfill \hbox{Men} &
\hfill \hbox{Women} & \hfill \hbox{Women} & 
\hfill \hbox{Men+women} & \hfill \hbox{Men+women}\cr
\hbox{}  & \hfill \hbox{} & \hfill \hbox{with} &
\hfill \hbox{} & \hfill \hbox{with} & 
\hfill \hbox{} & \hfill \hbox{with}\cr
\qtb
\hbox{}  &  \hfill \hbox{} & \hfill \hbox{disability} &
\hfill \hbox{} & \hfill \hbox{disability} & 
\hfill \hbox{} & \hfill \hbox{disability}\cr
\noalign{\hrule}
\qth
16 &  \hfill  10& \hfill 0& \hfill 28& \hfill 6& \hfill 38& \hfill 6\cr
20&  \hfill 42& \hfill 4& \hfill 128& \hfill 9&  \hfill 170& \hfill 13\cr
25&  \hfill 93& \hfill 5& \hfill 464& \hfill 21& \hfill 557& \hfill 26\cr
30&  \hfill 139& \hfill 4& \hfill 699& \hfill 30& \hfill 838&\hfill 34\cr
\qtb
35&  \hfill 209& \hfill 17& \hfill 1,005& \hfill 64& \hfill 1,214& \hfill 81\cr
\noalign{\hrule}
} $$
\vskip 1.5mm
\small
Notes: All numbers are for widowed persons.
``Disability'' refers to a condition which prevents
people from working. This ``work disability'' variable
was given in the censuses of 1980 and 1990. Needless to say,
the small number of cases in the 16-30 range gives rise to
strong inter-age statistical fluctuations.
\qL
{\it Source: 5\% sample of the census of 1980: Ruggles et al. 2017
(IPUMS).}
\vskip 5mm
\hrule
\vskip 0.7mm
\hrule
\end{table}
%%-----------------------------------------------

Two measures were taken to limit the impact of
statistical fluctuations arising from such small numbers.
\qee{1} The analysis used both the samples of 1980 and 1990 which 
together comprise 18 million individual records.
\qee{2} A centered moving window averaging was applied. 
We tried
widths of 5 and 11 years and the later proved the most satisfactory,

\qA{Uniformity test}

In order to make sure that the shape of the age-specific 
disability ratio is not brought about by a sub-sample of
outliers, we tested some 10 sub-samples each comprising
2 million records. They all led to curves of same shape
peaking in the 20-30 age interval and decreasing toward 1.2
at old ages.

\qI{Results}

\qA{Observations}

%
%%-----------------------------------------------
%%%% FIG -> DISABILITY RATIO ACCORDING TO MARITAL STATUS
\begin{figure}[htb]
\centerline{\psfig{width=17cm,figure=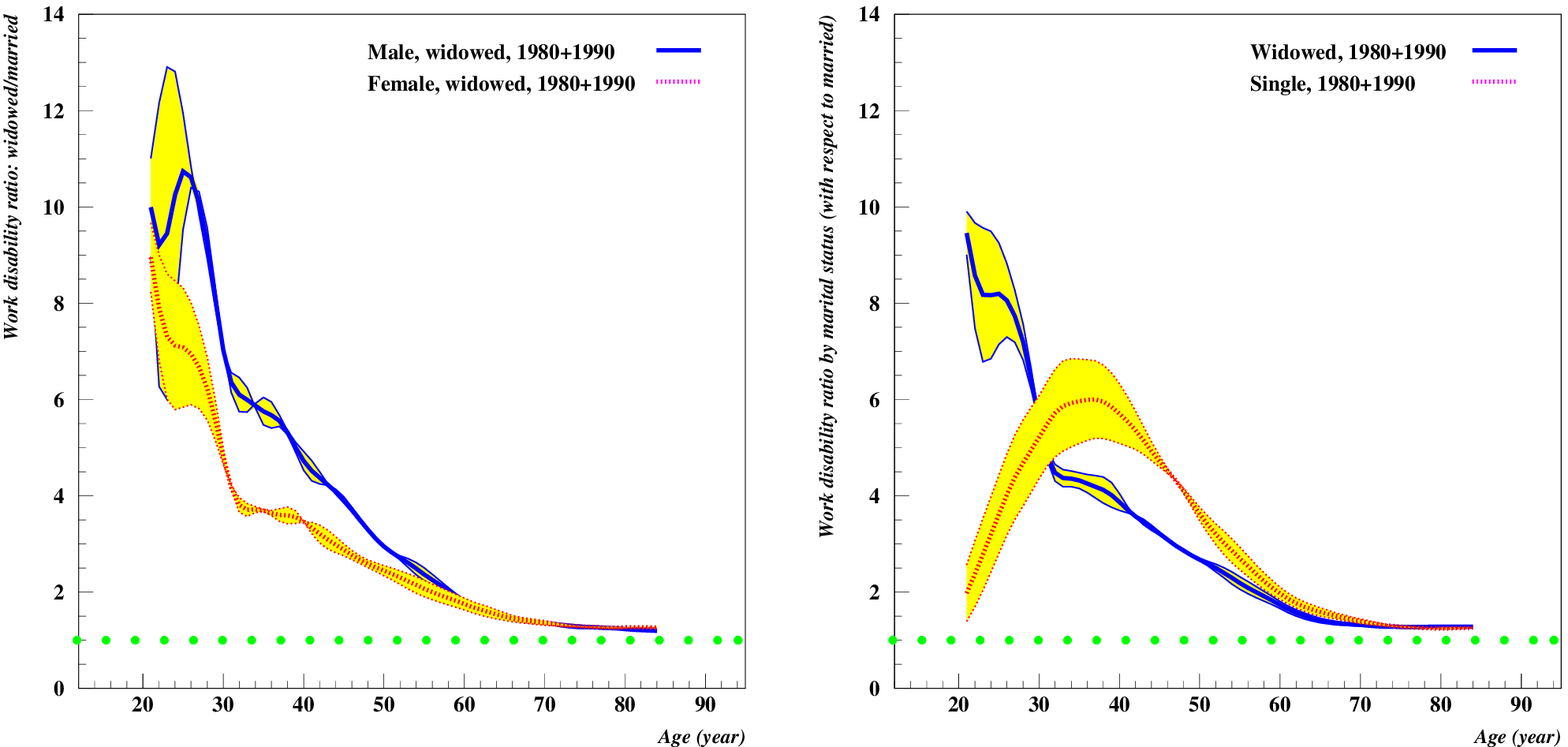}}
\qleg{Fig.\qhu 2a,b\qhv Age-specific disability ratio
with respect to married persons in the United States.} 
{(a) The disability ratio is the disability rate of widowed persons
divided by the disability rate of married persons. (b) The
curve for widowed persons is the average of the male and female
curves of (2a); ``single'' refers to persons who never got
married. Fig. 2b should be compared with Fig. 1 which is similar
except that disability is replaced by death.
Because there is a data point for each year, the 
error bars are not drawn as bars but are
shown as a (yellow) error band. It
represents $ \pm \sigma $ where $ \sigma $
is the standard deviation of the average.} 
{Source: 5\% samples of the US censuses of 1980 and 1990;
available from Ruggles et al. 2017 (IPUMS).}
\end{figure}
%-------------------------------------------------

The term ``disability ratio'' refers to the following variable:
$$ r(t;s)=f(t;s)/f(t;m) \hbox{ where: } t=\hbox{age},\ 
s=\hbox{marital status},\ m=\hbox{married} $$

$ f(t;s) $ and $ f(t;m) $ 
represent the fraction with disability in each group,
that is to say: 
$$ f(t;s)={
\hbox{Number of persons of marital status } s \hbox{ with a disability}
\over
\hbox{Total number of persons of marital status } s } $$

\qpar
Fig. 2a shows the disability ratios for
$ s=\hbox{widowed males} $ and $ s=\hbox{widowed females} $
respectively. \qL
Fig. 2b compares the disability ratios for
$ s=\hbox{widowed persons (male or female)} $ and
$ s=\hbox{single (i.e. never married) persons} $ respectively.\qL
These figures lead to the following conclusions.
\qbu Between the ages of 20 and 35 the curves for
widowed and single are very different: decreasing with
age for widowed and increasing for single persons. This observation
confirms what was found with Chinese death-ratio data.
For US data (Fig. 1) the difference was less striking
in the sense that in the age interval $ (20,30) $
the two curves are parallel.
\qbu Fig. 2a does not show any fundamental difference between
men and women except that the disability ratio of widows
is slightly lower than for widowers.
\qbu The peak for widowed persons reaches a level of about
10 which is intermediate between the value of 6 observed in the US
(Fig. 1) and the value of 20 observed in China.
\qpar

Incidentally, it should not come as a surprise that the shape
of the death ratio for young people
is country-dependent. This is due to the fact
that between 16 and 30 the main causes of death are not
diseases but external causes
such as traffic accidents, homicide or suicide. The frequency of
traffic accidents is of course conditioned by the number of
young people who drive cars or motorbikes.

\qA{Other groups of non-married persons}

So far we have considered only disability ratios for
widowed and single persons.
But the marital status variable of the census 
defines 6 different situations. Their definitions
and respective fractions in 1980 are as follows
(in 1990 the percentages are almost identical).\qL
(1) married, spouse present: 56\%,\qL
(2) married, spouse absent: 1.2\%, \qL
(3) separated: 2.2\%,  \qL
(4) divorced: 6.2\%, \qL
(5) widowed: 7.6\%, \qL
(6) single: 26\%.

Groups 2 and 3 are too small to be analyzed in a meaningful
way. Group 4, divorced persons, leads to a 
disability ratio curve
which is intermediate between ``widowed'' and ``single''. 
This contrasts
with ``widowed'' which has an rising part which peaks around the age
of 32 which is slightly earlier than the curve for ``single''.
The peak reaches a level of about 3, about 
2/3 the level reached by the curve for ``single'' and
only 1/3 the level reached by the curve for ``widowed''.

\qI{Health impact of a change in living place}

\qA{A testable prediction  of the ``Transient Shock'' conjecture}

For death occurrences the only data available
about the deceased are those recorded on the death certificate.
This includes only basic information such as
age, cause of death, marital status.
For the persons enumerated in a census much more
information is available which can be linked to the 
data about disability. Here this kind of linkage
is illustrated  by a particular interesting case which yields
a test of the ’transient shock” conjecture proposed in one of our earlier
papers.
%
%%-----------------------------------------------
%%%% FIG -> IMPACT OF A CHANGE OF LIVING PLACE ON DISABILITY
\begin{figure}[htb]
\centerline{\psfig{width=17cm,figure=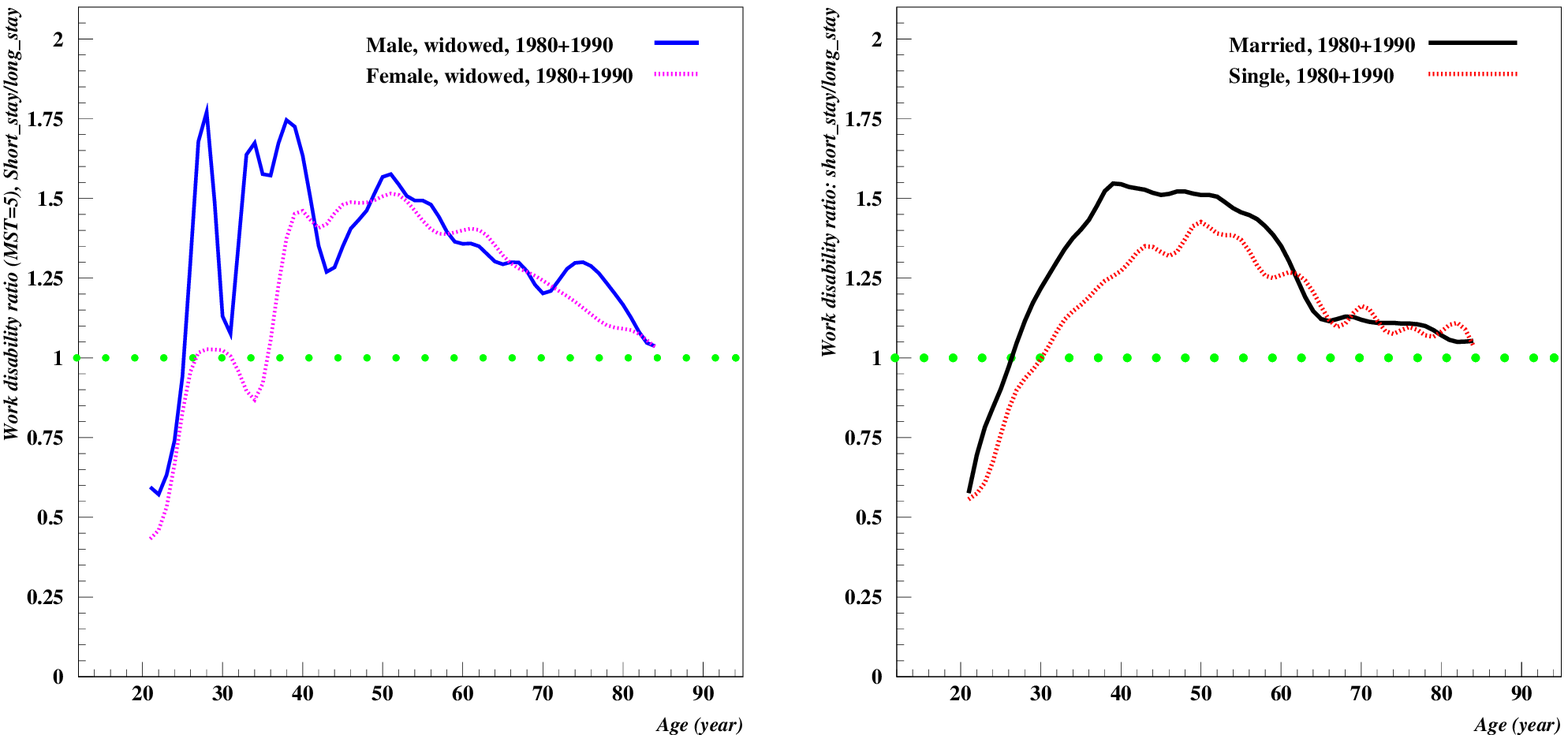}}
\qleg{Fig.\qhu 3a,b\qhv Age-specific disability ratio
according to time of residence in one and the same place.} 
{The disability ratio is the disability rate of
persons who have been moving to another place in the 18 months
preceding the census interview divided by the disability rate
of persons who have remained in the same place for more
than 18 months. For the sake of clarity the error bands have been
omitted; they have basically the same shape as in Fig. 2.} 
{Source: The results are based on the 18,037,222 records of 
the 5\% samples of the US censuses of 1980 and 1990;
available from Ruggles et al. 2017 (IPUMS).}
\end{figure}
%-------------------------------------------------
%
The ``Transient Shock'' conjecture introduced in Richmond et
al. (2016b) posits that:
\qdec{``Any abrupt change
in living conditions generates a mortality spike which acts as a kind
of selection process.''}
\qL
Moving from one place to another is a fairly sudden change although
hardly as dramatic as moving from home into
a nursing home%
\qfoot{An effect of that kind for
 elderly persons was studied in Richmond et al. 
(2016b).}%
.
This conjecture leads us to expect that a change of residence will have
an adverse effect on health; although because such
changes are fairly common one expects a small impact.
Yet thanks to the disability variable it is possible to test this
prediction.

\qA{Data and methodology}

In the censuses of 1980 and 1990 the question about the
living place was as follows.\qL
``When did the person move into this house or apartment?''\qL
Because it is well known that recollection of 
the date of past events
is fairly unreliable, the questionnaire proposed a number
of fairly broad time intervals. 
In 1980 the first interval was 1979--1980,
the second was 1975--1978 and the latter corresponded to
longer stays in the same place. Altogether there were 6 code
numbers with the final one standing for
``moved in more than 31 years ago''.
\qpar

Based on the analysis conducted in
Richmond et al. (2016b) we expect adverse effects to appear fairly
quickly after the change. This led us to consider only
two cases:  \qL
(i) Short stay: 1979--1980 
(ii) Long stay: all years before 1979.\qL
If we admit that the census question was asked in mid-1980,
the first case corresponds to a stay of between 0 month
(for a change occurring just prior to the census interview)
and 18 months (for a change occurring on 1 January 1979).
Thus, for this short-stay case the average length is 9 months.
\qpar

Because we expect the long-stay case to have a lower disability
rate than the short-stay case, it will be the analog of the
married status considered in the previous section.
Thus, in the same way as we computed disability ratios with
respect to the married status, here we will compute
disability ratios with respect to the long-stay case.
As before we compute this ratios for all ages between 16 and 89.
In addition we repeat this calculation for different
marital statuses.

\qA{Results} 

With respect to the analysis conducted in the previous section
adding a new variable namely the length of stay will further reduce
the number of persons who qualify. For that reason we 
need big samples and this led us to analyze the merged data
of both the 1980 and 1990 censuses.
\qpar
The curves presented 
in Fig. 3 were computed from the eighteen million
data lines of this merged file. 
They correspond to the following disability ratio:\qL
$ r(t;s)=f(t;short stay;s)/f(t;long stay;s) $, where $ t,s $ and $ f $
have the same meaning as previously. 
For ages over 25 the curves show indeed
disability ratios which are larger than one, thus confirming the
prediction based on the ``Transient Shock'' effect. 

\qA{Discussion}

The curves have also some surprising features. 
\qbu The maximum level which is reached between the ages of 40 and 60
is almost independent of the marital status.
\qbu After the age of 60 the curves fall until converging
towards a stationary level of 1.07.
\qpar

What is surprising in these observations can be summarized by saying
that one would expect the impact of relocation to be stronger
for more ``fragile'' groups.
It would be reasonable to think that widowed persons are more
fragile with respect to relocation than are married persons.
However, the results show basically the same effect for
widowed and married people.\qL
Similarly one would expect elderly people to be more fragile
than people in their 40s. However, the results of Fig. 3a,b show 
the opposite. They reveal a
relocation effect that is smaller for elderly groups
than for midlife people.
\qpar
%
%%-----------------------------------------------
%%%% FIG -> FRACTION OF DISABILITY: SHORT VERSUS LONG STAY
\begin{figure}[htb]
\centerline{\psfig{width=10cm,figure=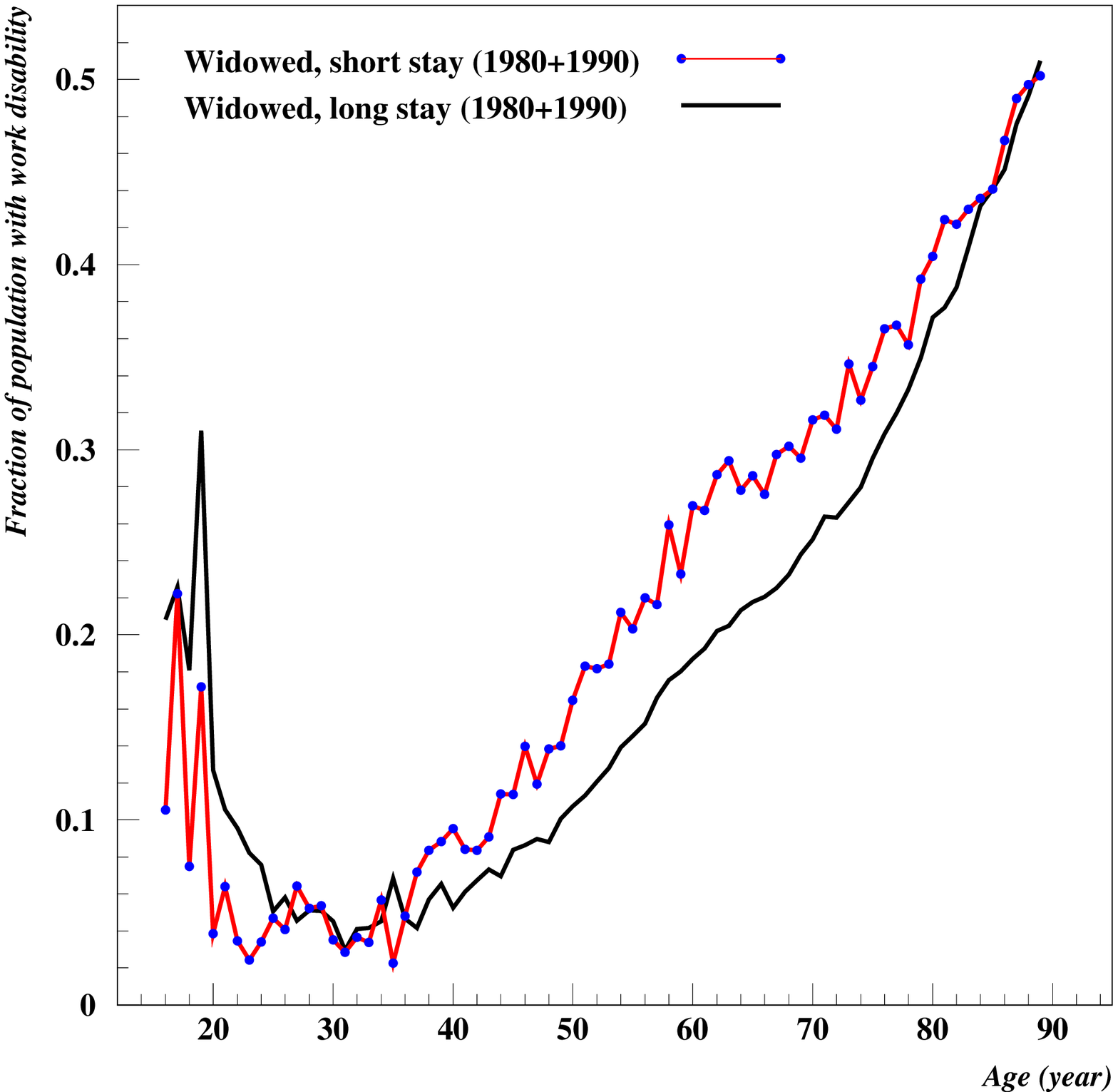}}
\qleg{Fig.\qhu 4\qhv Fractions of widowed groups with disability:
short stay versus long stay.} 
{``Short stay'' means same residence during a lapse of time
comprised between 0 and 18 months.
As both fractions increase along with age, the relocation
effect becomes drowned in the other many disability causes; as the
later are common to the two subgroups the impact of the
length of stay is reduced.} 
{Source: 5\% samples of the US censuses of 1980 and 1990;
available from Ruggles et al. 2017 (IPUMS).}
\end{figure}
%-------------------------------------------------
One can propose the following explanation.\qL
With increasing age come more and more disability factors mostly
related to health issues; relocation represents one of these factors
but as the number of the other factors increases the relative weight
of relocation decreases. This explanation is confirmed
by comparing the numerator and denominator of the disability ratio
(Fig. 4). As age increases, both terms increase 
but at the same time the gap between
the two curves narrows.
\qpar

In the next section we use the disability data to test
(or rather re-test)
a surprising effect already identified in 
Richmond et al. (2016b).

\qI{Short-term disability increase following marriage}

\qA{Method and data}

Based on the ``Transient Shock'' conjecture it was predicted
in Richmond et al. (2016b) that after marriage
there should be a temporary
increase in the mortality rate. This prediction was tested
and confirmed by three different methods which all
relied on mortality data. The fact that disability
rates can be used as a proxy for death rates opens
a new possibility and it is therefore interesting
to see whether our previous tests can be complemented.
\qpar
This investigation is based on the answers to the following
question which was asked in the 1\% censuses conducted between 2008
and 2015%
\qfoot{Actually as they are done on 1\% samples
they are not real full scale censuses. Such surveys 
are called ``American Community Surveys'' (ACS).}%
.
\qdec{``In the past 12 months did this person get married?''}
\qpar
If there is a temporary surge in disability in the months
following marriage, one should see an inflated rate for the persons
who got married within the past 12 months with respect to
those who have been married for a longer time. Fig. 5
shows that this is indeed the case except (for reasons as yet unknown)
for the youngest and oldest age intervals, namely 16-25 and 76-85.
%
%%-----------------------------------------------
%%%% FIG -> MARRIED IN PAST 12 MONTHS
\begin{figure}[htb]
\centerline{\psfig{width=10cm,figure=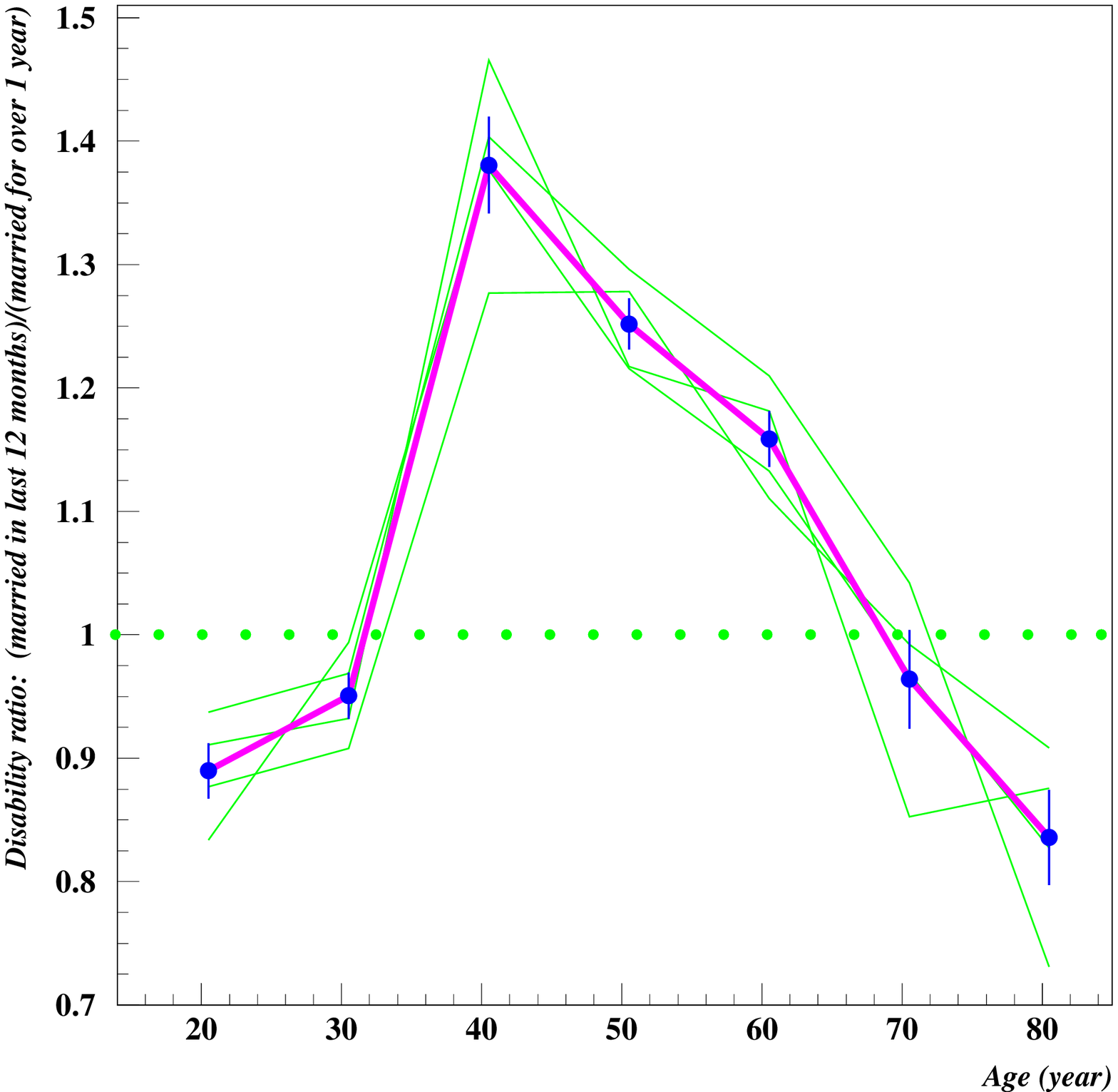}}
\qleg{Fig.\qhu 5\qhv Disability rate of persons who
got married in past 18 months divided by the disability rate of
persons who have been married for a longer time.} 
{The data points are for 10-year age intervals: 
$ (16-25), (26-35),\ldots , (75-85) $. 
The four thin lines in green
correspond to: (i) 2008-2010 (7,203,967 persons), (ii)
2011-2012 (4,988,147 persons), (iii) 2013-2014 (5,035,986 persons),
(iv) 2015 (2,542,244 persons).} 
{Sources: Data from the annual American Community Surveys from
2008 to 2015;
available from Ruggles et al. 2017 (IPUMS).}
\end{figure}
%-------------------------------------------------

\qA{Selection of the disability variable}

Note that for the ACS surveys
the disability variable is not defined
in the same way as in the censuses of 1980 and 1990.
Here, there are 3 different disability variables corresponding
to different aspects of the situation of the persons:
$ d_1 $: mobility at home, $ d_2 $: mobility outside home,
$ d_3 $: difficulty in bathing, dressing. It turns out that as
a function of age, $ \log(d_1),\ \log(d_2),\ \log(d_3) $ 
are highly correlated:
$ r(1,2)=0.93,\ r(1,3)=0.97,\ r(2,3)=0.99 $. With respect
to the mortality rate $ D $ in seven 10-year age intervals from 16 to 85 
they have the following relationships.\qL
$ d_1\sim D^{1.21},\ d_2\sim D^{1.93},\ d_3\sim D^{1.53} $.\qL
So we compute our results using $ d_1 $ as the condition
most closely connected with the mortality rate; the two variables
have a coefficient of correlation of $ 0.98 $.

\qA{Discussion}

The short term disability surge observed in the months
following marriage is nothing mysterious. It is a
direct  consequence of the fact that at same age single
persons have a higher disability rate than married persons.
Naturally, although the marriage itself
is instantaneous, the transition from condition 1 (single)
to condition 2 (married) is certainly {\it not} 
instantaneous. Thus, among the persons who responded that
they got married in the past 12 months, there are some
which still carry their former disability rate.
\qpar

The investigation done in Richmond et al. (2016b, Fig. 9b)
showed that in the age interval 30-40 the death rate
in the months following marriage was even higher than
the death rate of single persons. A comparison of Fig. 1b and
Fig. 5 shows that this does {\it not} hold for disability rates.
The reason of that discrepancy remains an open question.

\qI{Conclusion}

\qA{Overall results} 

In this paper we exploited the fact that disability rates
can be used as near-substitutes for mortality rates.
Whereas the second are based on death certificates, the
first are recorded individually in some (but not all) censuses.
This gives much more flexibility because censuses
record more personal information than that given
on death certificates.
\qpar

As a result we were were able to estimate the impact on
health of three different conditions: (i) marital
status (ii) moving from one living place to another
(iii) getting married.
For (i) and (iii) we had a prior knowledge of what
to expect from a previous study based on mortality
rates. Our observations led to effects similar
to what was seen with mortality rates thus confirming
that the disability rates are acceptable substitutes
for mortality rates. 
\qpar

The effect (ii) of a living place change in the 18 months
preceding the census interview
could be predicted based
on the ``Transient Shock'' conjecture. 
The results show that, on average over all ages, 
disability rates are inflated by a factor 1.5
with respect to the persons of same age and same marital status
who did not move.

\qA{Assets and promises of biodemography}

We finish with a word about the field of
biodemography. A few physicists have recently begun to explore this field
(see in particular Viswanathan et al. 2011), but
so far
it has not attracted their attention to the same degree as has finance.
When econophysics began some 20 years ago it mainly
focused on finance and in particular the study of stock prices.
Moving from stock markets to biodemography may seem
a big shift but unlike 200 hundred years ago when John Graunt and Edmund Halley were exploring similar data the two fields now share at least one
characteristic, namely the existence of broad and
massive data bases. \qL
As attested by our recent papers, such data sets
can now be readily interrogated in order to illuminate
regularities whose level of noise is much lower than that of
stock prices, not to speak of transaction volumes
which are even more volatile. \qL
Moreover, in biodemography
there are numerous well-defined and intriguing questions
that can be asked. By way of illustration
we note two: (i) Why are suicide rates
highest in May and lowest in December? (ii) Why are the death
rates of young widowers some 6 times higher than those for
married persons of the same age? At the moment the answers elude us.

\vskip 5mm

{\bf \large References}

\qparr
Bertillon (L.-A.) 1872: Article ``Mariage'' in: 
Dictionnaire Encyclop\'edique
des Sciences M\'edicales (Encyclopedic Dictionary of Medical
Sciences), vol. 5, 1872, pp. 7–52. 2nd series. \qL
[Available on ‘‘Gallica’’, the website
of digitized publications of the ``Biblioth\`eque Nationale de
France'' (French National Library) at:
http://www.bnf.fr].

\qparr
Richmond (P.), Roehner (B.M.) 2016a:
Effect of marital status on death rates. Part 1: High accuracy
exploration of the Farr-Bertillon effect.
Physica A 450,748–767.

\qparr
Richmond (P.), Roehner (B.M.) 2016b:
Effect of marital status on death rates. Part 2: Transient
mortality spikes.
Physica A 450,768–784.

\qparr
Ruggles (S.), Genadek (K.), Goeken (R.), 
Grover (J.), Sobek (M.) 2017
Integrated Public Use Microdata Series (IPUMS). 
University of Minnesota, Minneapolis (Minnesota).

\qparr
Viswanathan (G.M.), Luz (M.G.E. da), Raposo (E.P.R.), 
Stanley (H.E.) 2011: 
The physics of foraging.
An introduction to random searches and biological encounters.
Cambridge University Press, Cambridge.
\end{document}